\documentclass[%
 aip,
 amsmath,amssymb,
 reprint,%
]{revtex4-1}

\usepackage{graphicx}
\usepackage{dcolumn}
\usepackage{bm}

\usepackage[utf8]{inputenc}
\usepackage[T1]{fontenc}
\usepackage{mathptmx}
\usepackage{etoolbox}

\makeatletter
\def\@email#1#2{%
 \endgroup
 \patchcmd{\titleblock@produce}
  {\frontmatter@RRAPformat}
  {\frontmatter@RRAPformat{\produce@RRAP{*#1\href{mailto:#2}{#2}}}\frontmatter@RRAPformat}
  {}{}
}%
\makeatother
\begin{document}

\preprint{AIP/123-QED}

\title[]{Flow Structure near Three Phase Contact Line of Low-Contact-Angle Evaporating Droplets}
\author{Zhenying Wang}
\email{zhenying.wang@aero.kyushu-u.ac.jp}
 \affiliation{Department of Aeronautics and Astronautics, Kyushu University, Nishi-Ku, Motooka 744, Fukuoka 819-0395, Japan}%
\affiliation{
 International Institute for Carbon-Neutral Energy Research (WPI-I$^2$CNER), Kyushu University, Nishi-Ku, Motooka 744, Fukuoka 819-0395, Japan}%

\author{George Karapetsas}%
\affiliation{ Department of Chemical Engineering, Aristotle University of Thessaloniki, Thessaloniki 54124, Greece}%
 
\author{Prashant Valluri}%
\affiliation{ Institute of Multiscale Thermofluids, School of Engineering, University of Edinburgh, Edinburgh EH9 3JL, UK}%
\author{Chihiro Inoue}
 \affiliation{Department of Aeronautics and Astronautics, Kyushu University, Nishi-Ku, Motooka 744, Fukuoka 819-0395, Japan}%
 
\date{\today}

\begin{abstract}
Flow structure near three phase contact line (TPCL) of evaporating liquids plays a significant role in liquid wetting and dewetting, liquid film evaporation and boiling, etc. Despite the wide focus it receives, the interacting mechanisms therein remain elusive and in specific cases, controversial. Here, we reveal the profile of internal flow and elucidate the dominating mechanisms near TPCL of evaporating droplets, using mathematical modelling, microPIV, and infrared thermography. We indicate that for less volatile liquids such as butanol, the flow pattern is dominated by capillary flow. With increasing liquid volatility, e.g., alcohol, the effect of evaporation cooling, under conditions, induces interfacial temperature gradient with cold droplet apex and warm edge. The temperature gradient leads to Marangoni flow that competes with outwarding capillary flow, resulting in the reversal of interfacial flow and the formation of a stagnation point near TPCL. The spatiotemporal variations of capillary velocity and Marangoni velocity are further quantified by mathematically decomposing the tangential velocity of interfacial flow. The conclusions can serve as a theoretical base for explaining deposition patterns from colloidal suspensions, and can be utilized as a benchmark in analyzing more complex liquid systems.
\end{abstract}

\maketitle
The intersection between liquid, solid and vapor phases, i.e., three phase contact line (TPCL), exists whenever a finite volume of liquid gets contact with a solid surface\cite{de1985wetting,bonn2009wetting,brutin2018recent}. The dynamics and phase change near TPCL indicate great importance in many industrial processes. Specifically, in phase change thermal management, the liquid microlayer that forms in nucleate boiling exhibits concentrated high heat flux\cite{judd1976comprehensive,burevs2021modelling,salmean2022flow}. An efficient liquid supply to this region is critical for preventing dry-out and avoiding thermal degradation. In inkjet printing and quantum dot fabrication, the flow state inside drying droplets greatly affects the way of particle/crystal assembly \cite{haverinen2009inkjet,kong20143d,li2015coffee}. Especially, the flow state near TPCL decides the dynamics of contact line and the pathway of particle distribution \cite{huh1971hydrodynamic}.

Studies on flow pattern inside evaporating droplets have received wide focus so far\cite{deegan1997capillary,zang2019evaporation,wang2022wetting,gelderblom2022evaporation,jang2023volatility}. For the basic case of pure liquid droplets, the internal flow changes with the wetting state \cite{hu2005analysis,pan2013assessment}, the strength of evaporation mass flux\cite{girard2006evaporation}, and the induced interfacial stresses\cite{hu2006marangoni,hu2005analysismarangoni}. In the past decades, continuous research efforts have been paid to quantify the evaporation fluxes and droplet motion, with theoretical analysis \cite{lebedev1965special,hu2002evaporation,picknett1977evaporation}, numerical modeling\cite{hu2002evaporation,dunn2009strong}, and experimental measurements\cite{dunn2009strong,stauber2014lifetimes}. However, in comparison to the in-depth understanding of evaporation kinetics, far less is known on the details of flow patterns near droplet contact line\cite{wang2019contact}, where concentrated interactions take place between evaporation mass flux, capillary stress, and thermal Marangoni stress (natural consequence due to evaporation cooling)\cite{shiri2021thermal}; these effects considerably complicate the flow dynamics, and may lead to instabilities in that region\cite{kavehpour2002evaporatively,shen2020surface}. 

In this research, we visualize the flow pattern near TPCL of evaporating droplets with microparticle image velocimetry (microPIV). The governing mechanisms are further evaluated through detailed mathematical modelling based on the lubrication theory. In microPIV experiments, we utilize fluorescent micro tracers (fluorescent polymer microspheres: excitation/emission 468/508 nm, diameter 1.0 $\mu$m) to map the flow field inside deposited droplets of  $0.5 \pm 0.002\ \mu L$  (see experimental methodology in Supplemental Material). The droplet, sitting on transparent slide glass, is irradiated by a green laser sheet (468nm) and focused with 20x lens from the bottom. In order to account for the effect of volatility, we utilized 1-Butanol and 2-Propanol (IPA) as the test liquids, which have similar surface tensions ($\sigma_\mathrm{Butanol}=24.57\ N/m$, $\sigma_\mathrm{IPA}=23.0\ N/m$), latent heats ($L_\mathrm{Butanol}=584\ kJ/kg$, $L_\mathrm{IPA}=663\ kJ/kg$ ), and dynamic viscosities ($\mu_\mathrm{Butanol}=2.573\ mPa\cdot s$, $\mu_\mathrm{IPA}=2.012\ mPa\cdot s$), but different saturation vapor pressures ($p_\mathrm{sat,Butanol}=580\ Pa$, $p_\mathrm{sat,IPA}=4420\ Pa$). Results indicate that the droplet experiences a fast spreading stage right after its contact with the slide glass, i.e., the inertia regime ($R\propto t^{1/4}\sim t^{1/2}$)\cite{bird2008short,ding2007inertial,chen2013inertial} (Regime I). The flow state near TPCL in this stage is hard to trace due to the rapid motion of contact line and the narrow field of focus with 20x lens. As the droplet reaches a maximal radius, the contact line pins onto the slide glass and the flow pattern inside the droplet gets stable (Regime II). At the final stage when the droplet becomes very thin, a 'rush hour' in the flow field takes place - flow velocity increases dramatically- until the droplet fully dries out (Regime III)\cite{marin2011order}. In this research, we focus on Regime II with steady flow state and characterize the flow structure near TPCL by tracking the trajectory of the fluorescent particles. 

\begin{figure}[ht]
\includegraphics[width=8cm]{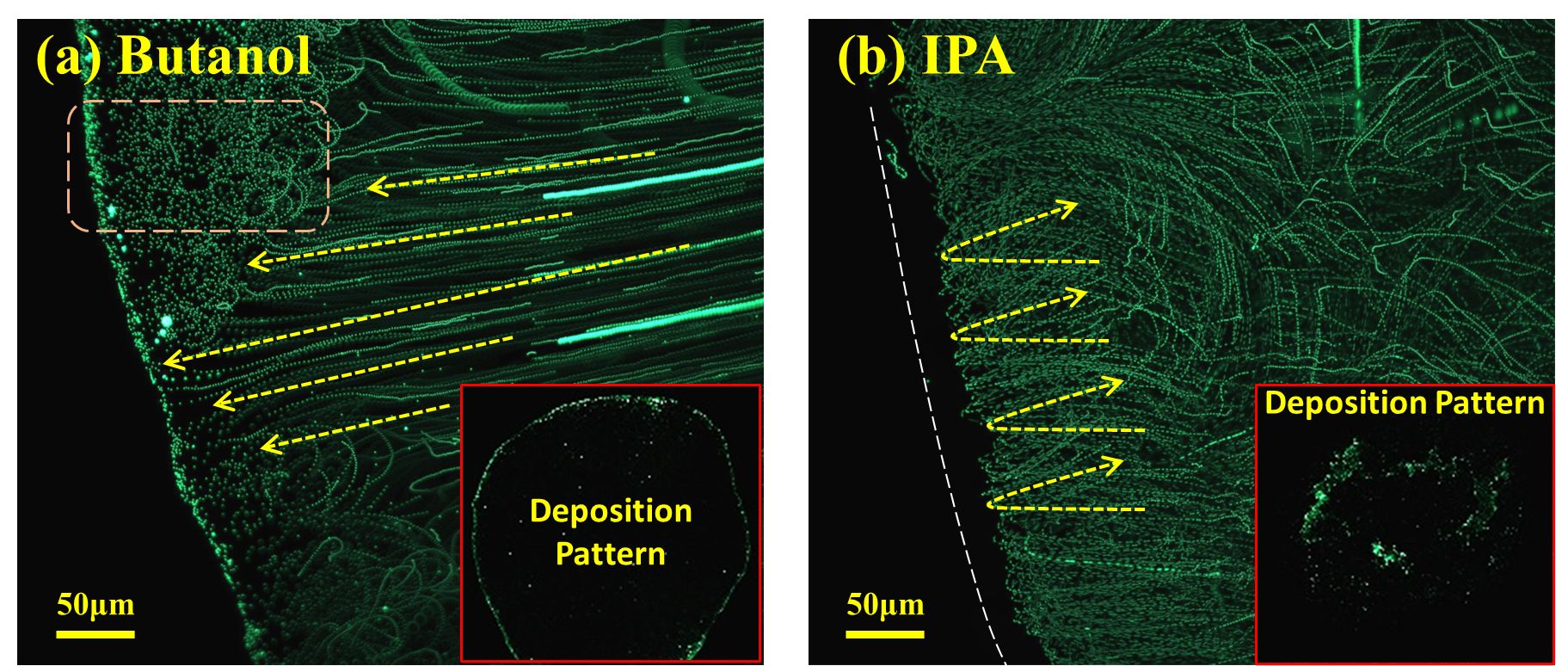}
\caption{\label{PIV_timelapse} Trajectory of tracing particles reveals the flow field near TPCL of evaporating drops. (a) An overall outwarding flow is observed along the bottom of a butanol droplet  - small disturbances exist while the particles ultimately move towards and get deposited near TPCL. (b) Flow near the bottom of an IPA droplet changes its direction at a position $\sim 20 \mu m$ from the TPCL, eliminating the coffee ring effect with more particles deposited in the central region.}
\end{figure}

Shown in Fig.~\ref{PIV_timelapse}, the tracer particles, in the case of butanol droplets, move continuously towards the contact line along the droplet bottom - most of these particles move all the way toward TPCL (marked as yellow arrows), with a few of them move slightly backward but finally return back to TPCL (circled with dashed box) - an indicative of weak instability in the azimuthal direction at the contact line as reported in previous studies\cite{cazabat2010evaporation}. The overall particle behaviors lead to the formation of so-called 'coffee ring' patterns as the droplet fully dries out, as marked with red frame in Fig.~\ref{PIV_timelapse}(a). But in the case of IPA, the tracer particles universally change their direction at a position around $20\sim 30\mu m$ from the TPCL as demonstrated by dot arrows in Fig.~\ref{PIV_timelapse}(b). The significant difference in the flow pattern by solely changing the liquid volatility drives us to further explore the interaction between phase change and fluid dynamics therein, in a quantitative way.

Specifically, we formulate a mathematical model for the problem of droplet evaporation on a thermally conductive substrate based on the lubrication theory \cite{wang2023spreading,wang2023intricate}. The lubrication approximation is valid for droplets with contact angles typically less than 40°, which is the case of this work in complete wetting and partial wetting scenarios using organic solvents on glass or metal substrates. The model takes account of the continuity, momentum and energy transport equations in the liquid phase, and heat conduction in the solid phase. At the liquid-gas interface, we derive the expression of interfacial mass flux by combining the Hertz-Knudsen equation\cite{plesset1976flow,moosman1980evaporating}, the chemical potential difference between liquid and gas\cite{atkins2014atkins}, and the ideal gas assumption. Other boundary equations include the jump energy balance, the stress balance and the kinematic boundary condition at the liquid-gas interface, as well as the continuity of heat flux and no-slip boundary condition at the liquid-solid interface. The force singularity that may arise from moving contact line and no-slip assumption is eliminated by assuming an ultrathin liquid film/precursor film existing at the solid surface in front of the TPCL\cite{ajaev2005spreading,wang2021dynamics}. See numerical methodology in Supplemental Materials.

\begin{figure}[ht]
\includegraphics[width=8.5cm]{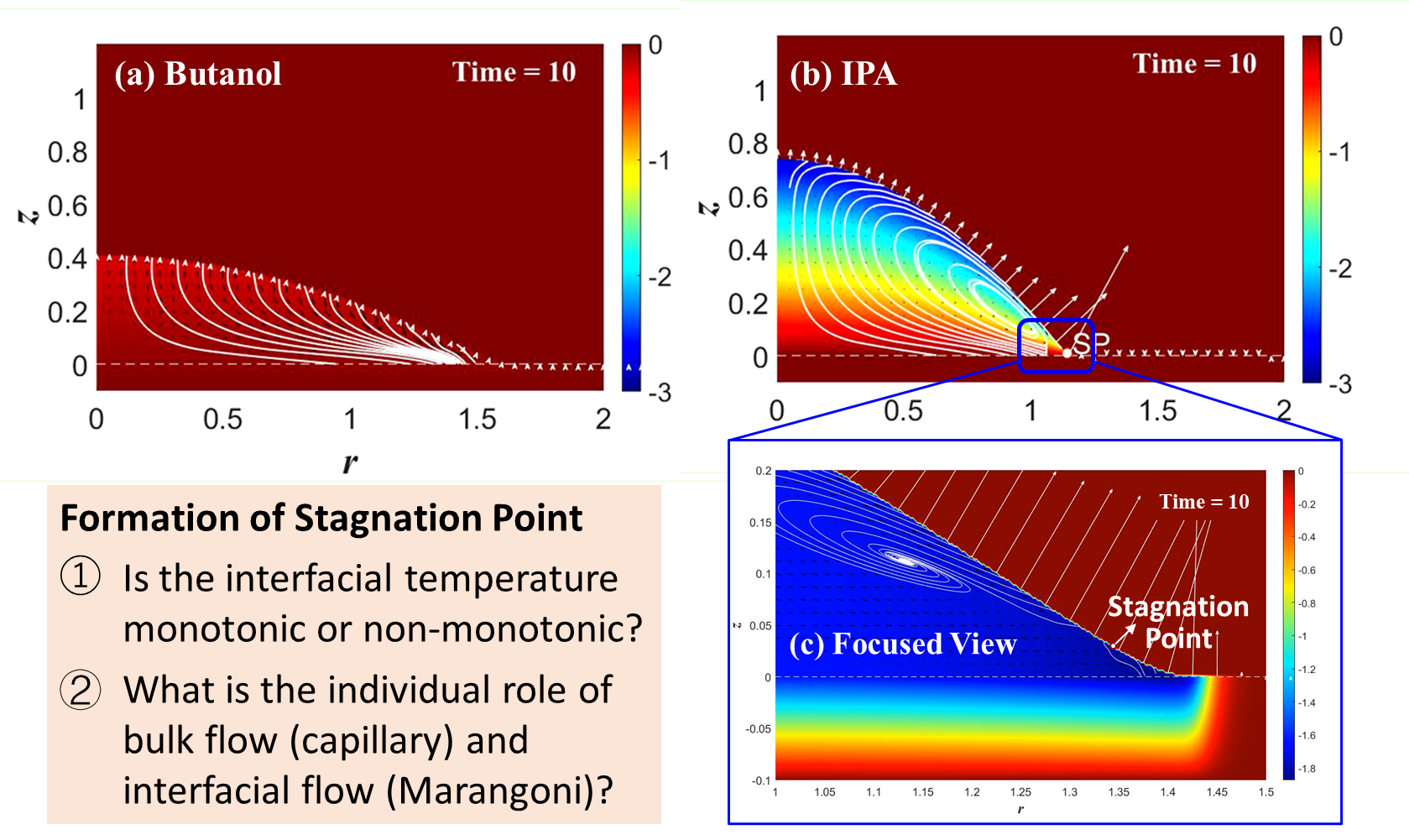}
\caption{\label{numerical_flow_pattern} Simulation results of the temperature and velocity fields inside evaporating droplets on highly conductive substrates, e.g. copper. (a) Butanol droplets show small temperature gradient and outwarding capillary flow; (b) IPA droplets indicate apparant temperature gradient due to evaporation cooling, and a convection vortex forms inside the droplet; (c) A close view to the contact line region reveals the formation of a stagnation point where the flow direction diverges.}
\end{figure}

As demonstrated by the side view of flow fields inside evaporating droplets at the steady stage (Fig.~\ref{numerical_flow_pattern}), the flow pattern in the case of low volatility butanol is dominated by an outwarding capillary flow (Fig.~\ref{numerical_flow_pattern} (a)). In comparison, flow inside a high volatility IPA droplet forms a convection vortex (Fig.~\ref{numerical_flow_pattern} (b)). Taking a close view on the contact line region of the IPA droplet (Fig.~\ref{numerical_flow_pattern}(c)), we can find a stagnation point with zero velocity that forms at the liquid-gas interface where the flow direction diverges. The position of the stagnation point locates at a dimensionless distance of $\sim$0.05 from TPCL with dimensionless contact radius $\sim$1.4  as in Fig.~\ref{numerical_flow_pattern}(c). This corresponds to a distance of $\sim35\ \mu m$ from TPCL with droplet contact radius $\sim 1\ mm$ , indicating quantitative correspondence with the microPIV results as shown in Fig.~\ref{PIV_timelapse} (bottom view) where the position is found to be $20\sim 30\mu m$ from the TPCL for IPA droplets with  $\sim 1\ mm$ contact radius.

There have been discussions on the formation mechanisms of the stagnation point over the years. Recent studies\cite{xu2007marangoni,ristenpart2007influence,xu2010criterion} have suggested that the temperature distribution at the droplet surface changes direction at different conditions, which results in the change of surface tension gradient and therefore the reversal of interfacial flow. In the work by Xu and Luo \cite{xu2007marangoni}, they calculated the temperature distribution at the liquid-air interface based on a static assumption with fixed droplet geometry and ignoring the convection by fluid flow inside the droplet. A maximum point in the interfacial temperature is found at a position $\sim10\mu m$ close to the TPCL. In the work of Ristenpart et al. \cite{ristenpart2007influence}, similarly, they considered a quasi-steady thermal transport process, and applied the Laplace's equation $\Delta T=0$ to derive the temperature field inside the droplet as the Peclet number (Pe, indicating the relative magnitudes of convective and conductive heat transfer) is small. The temperature gradient at the droplet surface, even though monotonic, is predicted to change with the ratio of liquid to solid thermal conductivity, and therefore the direction of interfacial flow changes. In subsequent studies\cite{xu2010criterion}, a criterion for flow reversal is proposed for droplets on substrates with finite thickness, where the flow reversal is attributed to the change of temperature gradient at the droplet surface.

However, according to available experimental results with infrared thermography in previous studies \cite{parsa2015effect,girard2008effect} as well as in some other numerical work \cite{dunn2009strong} including ours, the temperature gradient along the droplet surface (ignoring hydrothermal instability) is typically found to be monotonic with a cold droplet apex and a hot edge, including the cases that were predicted as in opposite direction based on the theory of Xu and Luo\cite{xu2010criterion}, and Ristenpart et al.\cite{ristenpart2007influence}, e.g. methanol and ethanol on PDMS, or water on glass of thickness 1 mm.

To address this inconsistency, we further checked the mathematical formulation and derivations in those studies that stand for the point of view that the change in temperature gradient leads to flow reversal inside an evaporating droplet. We find that, firstly, the mathematical modeling by Xu et al.\cite{xu2010criterion} and Ristenpart et al.\cite{ristenpart2007influence} mainly considered the thermal transport process with fixed droplet geometry, i.e., the energy equation and thermal boundary conditions with interfacial phase change. The lack of momentum equation results in the 'lack' of influence from bulk flow in their numerical/theoretical results, i.e., the capillary flow driven by Laplace pressure and the preferential evaporation at the contact line. Therefore, solely the Marangoni stress that arises from temperature gradient is accounted to explain the flow pattern inside the droplet. 

\begin{figure}[ht]
\centering
\includegraphics[width=9cm]{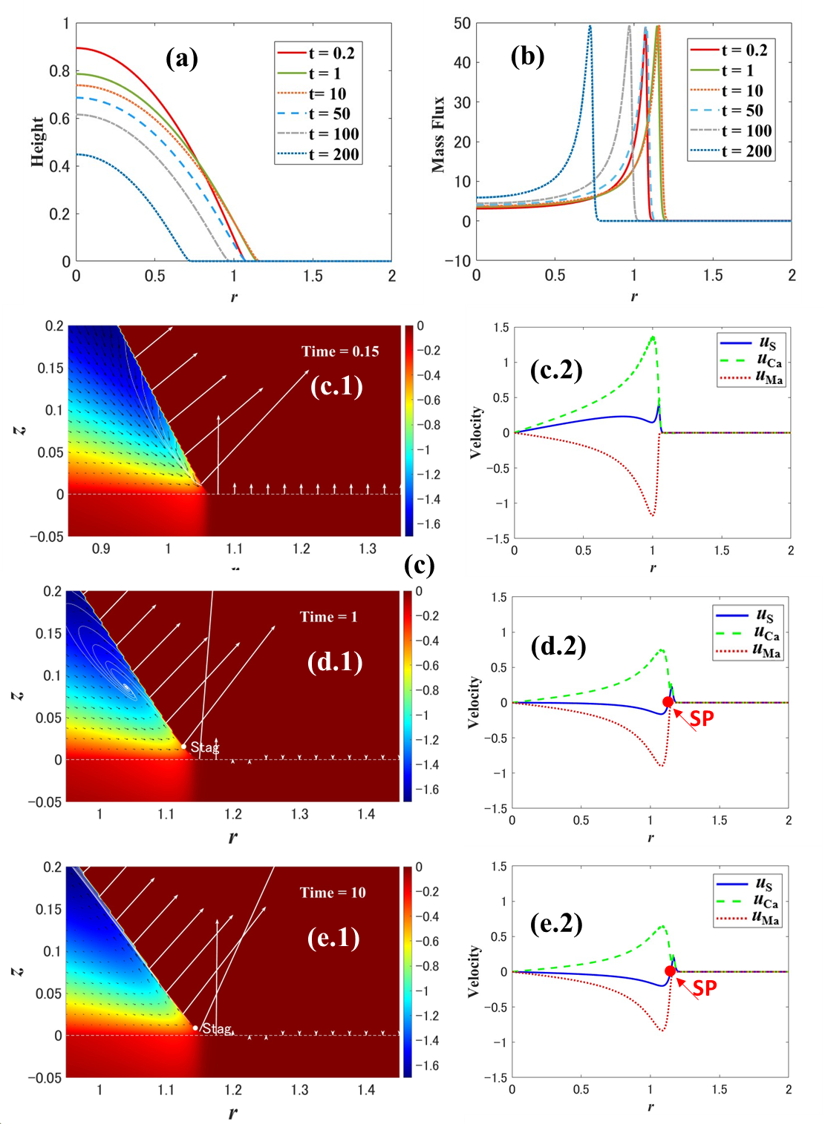}
\caption{\label{velocity_decomposition} Evolution of (a) droplet profile, (b) interfacial mass flux with dimensionless time $t$, and (c.1, d.1, e.1) Development of flow field inside an IPA droplet on a copper substrate along with (c.2, d.2, e.2) decomposition of interfacial velocity ($u_\mathrm{S}$: total tangential velocity; $u_\mathrm{Ca}$: capillary velocity; $u_\mathrm{Ma}$: Marangoni velocity) at corresponding moments.}
\end{figure}

Secondly, in their formulation of interfacial mass flux, the expression derived in the work of Deegan et al. \cite{deegan1997capillary} is utilized, which is basically derived from the analogy of vapor diffusion to an electrostatic field, expressed as $J(r)\propto (R-r)^{-\lambda},\ \lambda=(\pi-2\theta)/(2\pi-2\theta)$, where $J$ is the interfacial mass flux along radius $r$-axis, $R$ is the contact radius, $\theta$ is the contact angle. The expression is reliable for isothermal state when the evaporation mass flux is small, while the error can be large when the droplet becomes highly volatile and the effect of evaporation cooling becomes non-negligible. For example, the deviation in evaporation rate can reach $\sim 30\%$ with and without taking account of the evaporation cooling effect for an Acetone droplet with 1mm contact radius and 40$^\circ$ contact angle on PTFE substrate \cite{dunn2009strong}. By further taking account of the reduction of saturation vapor pressure due to evaporation cooling, Dunn et al.\cite{dunn2009strong} provided a better fitting to the evaporation flux. With modified expression of interfacial mass flux, the temperature distribution along the droplet surface is predicted as warmest near the contact line and coolest at the center of the droplet in all situations they considered, which corresponds with our model with Hertz-Knudsen type derivation of evaporation flux taking account of the surface curvature, the liquid-gas temperature difference, and the reduction in saturation vapor pressure due to evaporation cooling (mass flux shown in Fig.~\ref{velocity_decomposition}(b)).

\begin{figure}[ht]
\includegraphics[width=8.5cm]{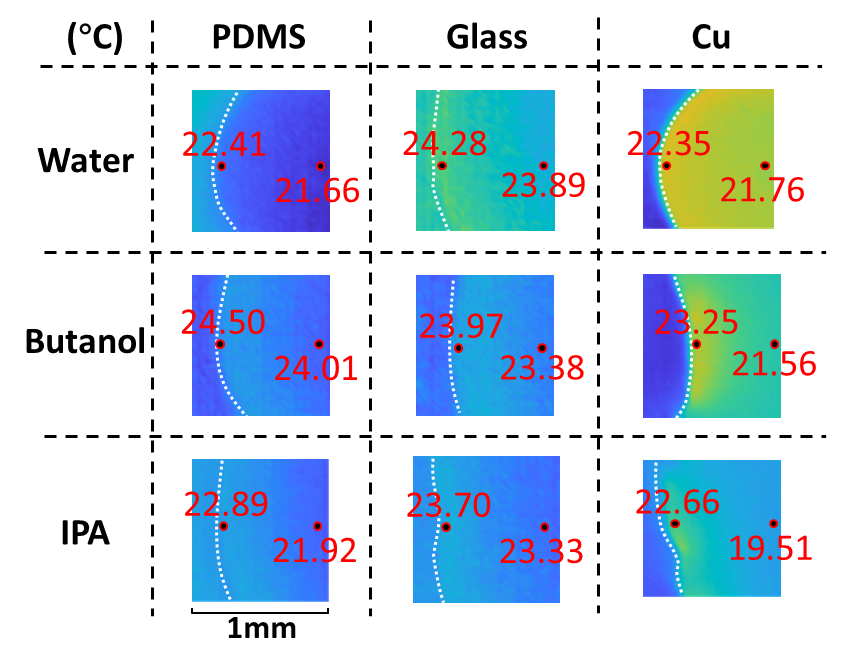}
\caption{\label{IRimages} Temperature field (top view) of evaporating droplets (water, butanol, IPA) on solid substrates (1mm PDMS, 1.1mm slide glass, 1mm copper) revealed by infrared thermography. Temperatures ($^\circ$C) at representative locations are marked in red.}
\end{figure}

Thirdly,  the thermal properties of liquids and solids are usually coupled (see representative examples in Supplementary Materials). For example, high surface tension arises from strong interactions between liquid molecules, which also makes the liquid less volatile. Substrates with low thermal conductivity usually have low surface energy, leading to large contact angles, e.g., polymer materials - this makes the interfacial mass flux near TPCL less divergent, and thus less apparent effect of evaporation cooling therein. Based on our experimental results with infrared camera as well as in available literature\cite{parsa2015effect,girard2008effect,josyula2018evaporation}, commonly-utilized liquids represented by butanol, water, and IPA, on solid substrates, e.g., copper, glass, PDMS, exhibit positive temperature gradient in radial direction, that is, warm edge and cold apex, as shown in Fig.~\ref{IRimages}.

In the experiments by IR imaging (Fig.~\ref{IRimages}), water droplets have contact angle of $30\pm5^\circ$ on glass, and $90\pm5^\circ$ on copper (with oxide layer) and PDMS. Butanol and IPA exhibit highly wetting
states with small contact angles, $5\sim10^\circ$, on glass, copper and PDMS. The experimental data of butanol and IPA provides comparable information on the effect of substrate conductivity and liquid volatility. Results show that both butanol and IPA exhibit large gradient of interfacial temperature on copper substrates and smaller on glass and PDMS. This is because on highly conductive substrates, the thermal resistance in the liquid phase becomes dominant in thermal transport, and therefore the temperature difference between droplet apex and edge becomes eminent. Additionally, on copper substrates, the temperature gradient in the case of IPA droplet is larger than that of butanol droplets due to stronger evaporation cooling effect ($P_\mathrm{sat,IPA,20^\circ C}=4420\ \rm Pa$, $P_\mathrm{sat,Butanol,20^\circ C}=580\ \rm Pa$). On less conductive PDMS substrates, the evaporation cooling effect weakens, but still the center part of the droplet has visibly lower interfacial temperature than the edge part. These experimental findings correspond with our numerical predictions on the effect of substrate conductivity\cite{wang2023intricate}.

Then the question comes that, if the temperature gradient at the liquid-air interface is positive with cold droplet apex and warm contact line in all cases that are investigated, then what causes the observed flow reversal and the formation of stagnation point at the droplet surface? This leads us to take a further look at the mathematical derivations, especially the expression of interfacial flow velocity, as well as its decomposition. 

In our mathematical model, the dimensionless form of surface velocity, $u_S$, derived from the momentum equation, is expressed as, $u_S=-\frac{h^2}{2\mu}\frac{\partial p}{\partial r}-\frac{h}{\mu}\frac{\partial T_S}{\partial r}$, where $h$ is the position of the liquid-air interface, $\mu$ is the dynamic viscosity of the liquid, $p$ is the liquid side pressure, and $T_S$ is the interfacial temperature. This can be further separated into a) capillary velocity, $u_{Ca}=-\frac{h^2}{2\mu}\frac{\partial p}{\partial r}$, where the height and shape of the liquid-gas interface (the latter affects the pressure gradient, $\frac{\partial p}{\partial r}$) play a decisive role, and b) Marangoni velocity, $u_{Ma}=-\frac{h}{\mu}\frac{\partial T_S}{\partial r}$, where the gradient of interfacial temperature, $\frac{\partial T_S}{\partial r}$, is the decisive factor. Taking a further view at the liquid side pressure, $p$, the expression is derived as $p=-\frac{\epsilon^2\sigma}{\mathrm{Ma}}(\frac{1}{r}\frac{\partial}{\partial r}(r\frac{\partial h}{\partial r})+\frac{1}{r^2}\frac{\partial^2h}{\partial\theta^2})-\frac{A}{h^3}$, where $\epsilon=\frac{H_0}{R_0}$ is the initial aspect ratio of the droplet, $\sigma$ is the surface tension, Ma is the Marangoni number, demonstrating the strength of Marangoni flow, and $A$ is the Hamaker constant, indicating the strength of van der Waals interaction. Here the van der Waals force, $\frac{A}{h^3}$,  only becomes significant in the precursor film region when the thickness of liquid film is down to the nanometer (nm) scale, and is negligible at the macroscopic and transitional regions with mm to $\mu m$ scale film thickness. The term $(\frac{1}{r}\frac{\partial}{\partial r}(r\frac{\partial h}{\partial r})+\frac{1}{r^2}\frac{\partial^2h}{\partial\theta^2})$ in the expression of $p$ indicates the surface curvature, i.e., geometry of the liquid-air interface. Due to the existence of interfacial tension $\sigma$, the surface curvature generates Laplace pressure, and therefore out-warding capillary flow, which tends to flatten the droplet. The gradient of interfacial temperature, $\frac{\partial T_S}{\partial r}$, as analyzed before, is positive in the $r$-direction, which generate inward flow along the liquid-air interface (from contact line to droplet center), opposite to the direction of capillary flow. 

With changing droplet geometry (e.g., spreading, receding) (Fig.~\ref{velocity_decomposition}(a)) and spatiotemporally varying mass flux (Fig.~\ref{velocity_decomposition}(b)), the relative strength of capillary flow and Marangoni flow changes with time. At the initial stage after a droplet is deposited onto a solid surface, the capillary effect dominates and induces internal flow directing from the droplet apex to the edge (shown by the flow field and the decomposition of surface velocity at dimensionless time $t$ = 0.15 in Fig.~\ref{velocity_decomposition}(c.1) and (c.2)), driving fast droplet spreading. As temperature gradient establishes across the droplet due to evaporation cooling, the Marangoni flow enhances with time as shown by the transition of flow pattern from Fig.~\ref{velocity_decomposition}(c.1) to Fig.~\ref{velocity_decomposition}(e.1), with corresponding decomposition of interfacial velocity as shown in Fig.~\ref{velocity_decomposition}(d.2) and (e.2). 

\begin{figure}[ht]
\centering
\includegraphics[width=9cm]{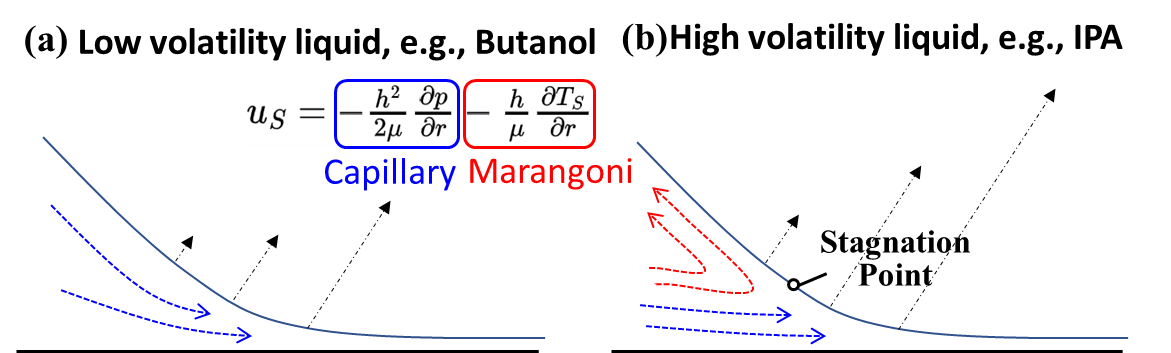}
\caption{\label{schematic_mechanism} Schematic of flow structure and the dominating mechanisms near TPCL of (a) low volatility liquids, e.g. butanol, and (b) high volatility liquids, e.g. IPA.}
\end{figure}

For liquids with low volatility, the Marangoni effect is not strong enough to compete with the capillary effect, and the flow direction is overall outwarding as demonstared by Fig.~\ref{schematic_mechanism}(a). For highly volatile droplets on thermally conductive substrates, the mechanisms can be described by Fig.~\ref{schematic_mechanism} (b). In such cases, the effect of evaporation cooling can be strong enough to generate Marangoni flow that overweighs the outward capillary flow near TPCL, which causes the formation of stagnation point ($u_\mathrm{S}=0$) where the flow direction diverges. 

For suspended particles in a volatile droplet such as alcohol, those that are transported to a position below the stagnation point will get to deposit near the contact line as the liquid film dries to a critical thickness (comparable to the particle size) due to preferential evaporation. Particles that are transported to a position higher than the stagnation point will be recirculated to the center area and move back again for reallocation. Therefore, the deposition pattern does not only depend on the capillary flow and the strength of thermal Marangoni flow (effect of evaporation cooling), but also relies on the motion of contact line. For achieving uniform deposition patterns from single component droplets, it is necessary to achieve smooth contact line receding that matches the deposition speed of the suspended particles\cite{li2020evaporating}. In view of the difficulty in manipulating the contact line motion, it is an easier approach to utilize binary solvents or surfactants so as to induce solutal Marangoni stresses that further complex the internal flow and strengthen the particle circulation. In the fabrication of 2D crystals, perovskite solar cells, and QLEDs, both solutal Marangoni effect and controllable contact line geometry have been utilized for generating uniform thin film coatings and enhancing the efficiency of energy conversion \cite{zhang2010colloidal,park2020scalable,kim20223d,gao2020printable,yu2020efficient}.

\setlength{\parskip}{1em}
The authors gratefully acknowledge the support received from Japanese Society for the Promotion of Science (JSPS), and ThermaSMART project of European Commission (Grant No. EC-H2020-RISE-ThermaSMART-778104). ZW and CI acknowledge the support from JSPS KAKENHI (Grant Nos. JP21K14097, JP21H01251, JP22K18771). 

\bibliography{References}

\begin{thebibliography}{53}%
\makeatletter
\providecommand \@ifxundefined [1]{%
 \@ifx{#1\undefined}
}%
\providecommand \@ifnum [1]{%
 \ifnum #1\expandafter \@firstoftwo
 \else \expandafter \@secondoftwo
 \fi
}%
\providecommand \@ifx [1]{%
 \ifx #1\expandafter \@firstoftwo
 \else \expandafter \@secondoftwo
 \fi
}%
\providecommand \natexlab [1]{#1}%
\providecommand \enquote  [1]{``#1''}%
\providecommand \bibnamefont  [1]{#1}%
\providecommand \bibfnamefont [1]{#1}%
\providecommand \citenamefont [1]{#1}%
\providecommand \href@noop [0]{\@secondoftwo}%
\providecommand \href [0]{\begingroup \@sanitize@url \@href}%
\providecommand \@href[1]{\@@startlink{#1}\@@href}%
\providecommand \@@href[1]{\endgroup#1\@@endlink}%
\providecommand \@sanitize@url [0]{\catcode `\\12\catcode `\$12\catcode `\&12\catcode `\#12\catcode `\^12\catcode `\_12\catcode `\%12\relax}%
\providecommand \@@startlink[1]{}%
\providecommand \@@endlink[0]{}%
\providecommand \url  [0]{\begingroup\@sanitize@url \@url }%
\providecommand \@url [1]{\endgroup\@href {#1}{\urlprefix }}%
\providecommand \urlprefix  [0]{URL }%
\providecommand \Eprint [0]{\href }%
\providecommand \doibase [0]{http://dx.doi.org/}%
\providecommand \selectlanguage [0]{\@gobble}%
\providecommand \bibinfo  [0]{\@secondoftwo}%
\providecommand \bibfield  [0]{\@secondoftwo}%
\providecommand \translation [1]{[#1]}%
\providecommand \BibitemOpen [0]{}%
\providecommand \bibitemStop [0]{}%
\providecommand \bibitemNoStop [0]{.\EOS\space}%
\providecommand \EOS [0]{\spacefactor3000\relax}%
\providecommand \BibitemShut  [1]{\csname bibitem#1\endcsname}%
\let\auto@bib@innerbib\@empty
\bibitem [{\citenamefont {De~Gennes}(1985)}]{de1985wetting}%
  \BibitemOpen
  \bibfield  {author} {\bibinfo {author} {\bibfnamefont {P.-G.}\ \bibnamefont {De~Gennes}},\ }\bibfield  {title} {\enquote {\bibinfo {title} {Wetting: statics and dynamics},}\ }\href@noop {} {\bibfield  {journal} {\bibinfo  {journal} {Reviews of modern physics}\ }\textbf {\bibinfo {volume} {57}},\ \bibinfo {pages} {827} (\bibinfo {year} {1985})}\BibitemShut {NoStop}%
\bibitem [{\citenamefont {Bonn}\ \emph {et~al.}(2009)\citenamefont {Bonn}, \citenamefont {Eggers}, \citenamefont {Indekeu}, \citenamefont {Meunier},\ and\ \citenamefont {Rolley}}]{bonn2009wetting}%
  \BibitemOpen
  \bibfield  {author} {\bibinfo {author} {\bibfnamefont {D.}~\bibnamefont {Bonn}}, \bibinfo {author} {\bibfnamefont {J.}~\bibnamefont {Eggers}}, \bibinfo {author} {\bibfnamefont {J.}~\bibnamefont {Indekeu}}, \bibinfo {author} {\bibfnamefont {J.}~\bibnamefont {Meunier}}, \ and\ \bibinfo {author} {\bibfnamefont {E.}~\bibnamefont {Rolley}},\ }\bibfield  {title} {\enquote {\bibinfo {title} {Wetting and spreading},}\ }\href@noop {} {\bibfield  {journal} {\bibinfo  {journal} {Reviews of modern physics}\ }\textbf {\bibinfo {volume} {81}},\ \bibinfo {pages} {739} (\bibinfo {year} {2009})}\BibitemShut {NoStop}%
\bibitem [{\citenamefont {Brutin}\ and\ \citenamefont {Starov}(2018)}]{brutin2018recent}%
  \BibitemOpen
  \bibfield  {author} {\bibinfo {author} {\bibfnamefont {D.}~\bibnamefont {Brutin}}\ and\ \bibinfo {author} {\bibfnamefont {V.}~\bibnamefont {Starov}},\ }\bibfield  {title} {\enquote {\bibinfo {title} {Recent advances in droplet wetting and evaporation},}\ }\href@noop {} {\bibfield  {journal} {\bibinfo  {journal} {Chemical Society Reviews}\ }\textbf {\bibinfo {volume} {47}},\ \bibinfo {pages} {558--585} (\bibinfo {year} {2018})}\BibitemShut {NoStop}%
\bibitem [{\citenamefont {Judd}\ and\ \citenamefont {Hwang}(1976)}]{judd1976comprehensive}%
  \BibitemOpen
  \bibfield  {author} {\bibinfo {author} {\bibfnamefont {R.}~\bibnamefont {Judd}}\ and\ \bibinfo {author} {\bibfnamefont {K.}~\bibnamefont {Hwang}},\ }\bibfield  {title} {\enquote {\bibinfo {title} {A comprehensive model for nucleate pool boiling heat transfer including microlayer evaporation},}\ }\href@noop {} {\  (\bibinfo {year} {1976})}\BibitemShut {NoStop}%
\bibitem [{\citenamefont {Bure{\v{s}}}\ and\ \citenamefont {Sato}(2021)}]{burevs2021modelling}%
  \BibitemOpen
  \bibfield  {author} {\bibinfo {author} {\bibfnamefont {L.}~\bibnamefont {Bure{\v{s}}}}\ and\ \bibinfo {author} {\bibfnamefont {Y.}~\bibnamefont {Sato}},\ }\bibfield  {title} {\enquote {\bibinfo {title} {On the modelling of the transition between contact-line and microlayer evaporation regimes in nucleate boiling},}\ }\href@noop {} {\bibfield  {journal} {\bibinfo  {journal} {Journal of Fluid Mechanics}\ }\textbf {\bibinfo {volume} {916}},\ \bibinfo {pages} {A53} (\bibinfo {year} {2021})}\BibitemShut {NoStop}%
\bibitem [{\citenamefont {Salmean}\ and\ \citenamefont {Qiu}(2022)}]{salmean2022flow}%
  \BibitemOpen
  \bibfield  {author} {\bibinfo {author} {\bibfnamefont {C.}~\bibnamefont {Salmean}}\ and\ \bibinfo {author} {\bibfnamefont {H.}~\bibnamefont {Qiu}},\ }\bibfield  {title} {\enquote {\bibinfo {title} {Flow boiling enhancement using three-dimensional contact-line pinning on hierarchical superbiphilic micro/nanostructures},}\ }\href@noop {} {\bibfield  {journal} {\bibinfo  {journal} {Nano Letters}\ }\textbf {\bibinfo {volume} {22}},\ \bibinfo {pages} {8487--8494} (\bibinfo {year} {2022})}\BibitemShut {NoStop}%
\bibitem [{\citenamefont {Haverinen}, \citenamefont {Myllyl{\"a}},\ and\ \citenamefont {Jabbour}(2009)}]{haverinen2009inkjet}%
  \BibitemOpen
  \bibfield  {author} {\bibinfo {author} {\bibfnamefont {H.~M.}\ \bibnamefont {Haverinen}}, \bibinfo {author} {\bibfnamefont {R.~A.}\ \bibnamefont {Myllyl{\"a}}}, \ and\ \bibinfo {author} {\bibfnamefont {G.~E.}\ \bibnamefont {Jabbour}},\ }\bibfield  {title} {\enquote {\bibinfo {title} {Inkjet printing of light emitting quantum dots},}\ }\href@noop {} {\bibfield  {journal} {\bibinfo  {journal} {Applied Physics Letters}\ }\textbf {\bibinfo {volume} {94}} (\bibinfo {year} {2009})}\BibitemShut {NoStop}%
\bibitem [{\citenamefont {Kong}\ \emph {et~al.}(2014)\citenamefont {Kong}, \citenamefont {Tamargo}, \citenamefont {Kim}, \citenamefont {Johnson}, \citenamefont {Gupta}, \citenamefont {Koh}, \citenamefont {Chin}, \citenamefont {Steingart}, \citenamefont {Rand},\ and\ \citenamefont {McAlpine}}]{kong20143d}%
  \BibitemOpen
  \bibfield  {author} {\bibinfo {author} {\bibfnamefont {Y.~L.}\ \bibnamefont {Kong}}, \bibinfo {author} {\bibfnamefont {I.~A.}\ \bibnamefont {Tamargo}}, \bibinfo {author} {\bibfnamefont {H.}~\bibnamefont {Kim}}, \bibinfo {author} {\bibfnamefont {B.~N.}\ \bibnamefont {Johnson}}, \bibinfo {author} {\bibfnamefont {M.~K.}\ \bibnamefont {Gupta}}, \bibinfo {author} {\bibfnamefont {T.-W.}\ \bibnamefont {Koh}}, \bibinfo {author} {\bibfnamefont {H.-A.}\ \bibnamefont {Chin}}, \bibinfo {author} {\bibfnamefont {D.~A.}\ \bibnamefont {Steingart}}, \bibinfo {author} {\bibfnamefont {B.~P.}\ \bibnamefont {Rand}}, \ and\ \bibinfo {author} {\bibfnamefont {M.~C.}\ \bibnamefont {McAlpine}},\ }\bibfield  {title} {\enquote {\bibinfo {title} {3d printed quantum dot light-emitting diodes},}\ }\href@noop {} {\bibfield  {journal} {\bibinfo  {journal} {Nano letters}\ }\textbf {\bibinfo {volume} {14}},\ \bibinfo {pages} {7017--7023} (\bibinfo {year} {2014})}\BibitemShut {NoStop}%
\bibitem [{\citenamefont {Li}\ \emph {et~al.}(2015)\citenamefont {Li}, \citenamefont {Lv}, \citenamefont {Li}, \citenamefont {Qu{\'e}r{\'e}},\ and\ \citenamefont {Zheng}}]{li2015coffee}%
  \BibitemOpen
  \bibfield  {author} {\bibinfo {author} {\bibfnamefont {Y.}~\bibnamefont {Li}}, \bibinfo {author} {\bibfnamefont {C.}~\bibnamefont {Lv}}, \bibinfo {author} {\bibfnamefont {Z.}~\bibnamefont {Li}}, \bibinfo {author} {\bibfnamefont {D.}~\bibnamefont {Qu{\'e}r{\'e}}}, \ and\ \bibinfo {author} {\bibfnamefont {Q.}~\bibnamefont {Zheng}},\ }\bibfield  {title} {\enquote {\bibinfo {title} {From coffee rings to coffee eyes},}\ }\href@noop {} {\bibfield  {journal} {\bibinfo  {journal} {Soft matter}\ }\textbf {\bibinfo {volume} {11}},\ \bibinfo {pages} {4669--4673} (\bibinfo {year} {2015})}\BibitemShut {NoStop}%
\bibitem [{\citenamefont {Huh}\ and\ \citenamefont {Scriven}(1971)}]{huh1971hydrodynamic}%
  \BibitemOpen
  \bibfield  {author} {\bibinfo {author} {\bibfnamefont {C.}~\bibnamefont {Huh}}\ and\ \bibinfo {author} {\bibfnamefont {L.~E.}\ \bibnamefont {Scriven}},\ }\bibfield  {title} {\enquote {\bibinfo {title} {Hydrodynamic model of steady movement of a solid/liquid/fluid contact line},}\ }\href@noop {} {\bibfield  {journal} {\bibinfo  {journal} {Journal of colloid and interface science}\ }\textbf {\bibinfo {volume} {35}},\ \bibinfo {pages} {85--101} (\bibinfo {year} {1971})}\BibitemShut {NoStop}%
\bibitem [{\citenamefont {Deegan}\ \emph {et~al.}(1997)\citenamefont {Deegan}, \citenamefont {Bakajin}, \citenamefont {Dupont}, \citenamefont {Huber}, \citenamefont {Nagel},\ and\ \citenamefont {Witten}}]{deegan1997capillary}%
  \BibitemOpen
  \bibfield  {author} {\bibinfo {author} {\bibfnamefont {R.~D.}\ \bibnamefont {Deegan}}, \bibinfo {author} {\bibfnamefont {O.}~\bibnamefont {Bakajin}}, \bibinfo {author} {\bibfnamefont {T.~F.}\ \bibnamefont {Dupont}}, \bibinfo {author} {\bibfnamefont {G.}~\bibnamefont {Huber}}, \bibinfo {author} {\bibfnamefont {S.~R.}\ \bibnamefont {Nagel}}, \ and\ \bibinfo {author} {\bibfnamefont {T.~A.}\ \bibnamefont {Witten}},\ }\bibfield  {title} {\enquote {\bibinfo {title} {Capillary flow as the cause of ring stains from dried liquid drops},}\ }\href@noop {} {\bibfield  {journal} {\bibinfo  {journal} {Nature}\ }\textbf {\bibinfo {volume} {389}},\ \bibinfo {pages} {827--829} (\bibinfo {year} {1997})}\BibitemShut {NoStop}%
\bibitem [{\citenamefont {Zang}\ \emph {et~al.}(2019)\citenamefont {Zang}, \citenamefont {Tarafdar}, \citenamefont {Tarasevich}, \citenamefont {Choudhury},\ and\ \citenamefont {Dutta}}]{zang2019evaporation}%
  \BibitemOpen
  \bibfield  {author} {\bibinfo {author} {\bibfnamefont {D.}~\bibnamefont {Zang}}, \bibinfo {author} {\bibfnamefont {S.}~\bibnamefont {Tarafdar}}, \bibinfo {author} {\bibfnamefont {Y.~Y.}\ \bibnamefont {Tarasevich}}, \bibinfo {author} {\bibfnamefont {M.~D.}\ \bibnamefont {Choudhury}}, \ and\ \bibinfo {author} {\bibfnamefont {T.}~\bibnamefont {Dutta}},\ }\bibfield  {title} {\enquote {\bibinfo {title} {Evaporation of a droplet: From physics to applications},}\ }\href@noop {} {\bibfield  {journal} {\bibinfo  {journal} {Physics Reports}\ }\textbf {\bibinfo {volume} {804}},\ \bibinfo {pages} {1--56} (\bibinfo {year} {2019})}\BibitemShut {NoStop}%
\bibitem [{\citenamefont {Wang}\ \emph {et~al.}(2022)\citenamefont {Wang}, \citenamefont {Orejon}, \citenamefont {Takata},\ and\ \citenamefont {Sefiane}}]{wang2022wetting}%
  \BibitemOpen
  \bibfield  {author} {\bibinfo {author} {\bibfnamefont {Z.}~\bibnamefont {Wang}}, \bibinfo {author} {\bibfnamefont {D.}~\bibnamefont {Orejon}}, \bibinfo {author} {\bibfnamefont {Y.}~\bibnamefont {Takata}}, \ and\ \bibinfo {author} {\bibfnamefont {K.}~\bibnamefont {Sefiane}},\ }\bibfield  {title} {\enquote {\bibinfo {title} {Wetting and evaporation of multicomponent droplets},}\ }\href@noop {} {\bibfield  {journal} {\bibinfo  {journal} {Physics Reports}\ }\textbf {\bibinfo {volume} {960}},\ \bibinfo {pages} {1--37} (\bibinfo {year} {2022})}\BibitemShut {NoStop}%
\bibitem [{\citenamefont {Gelderblom}, \citenamefont {Diddens},\ and\ \citenamefont {Marin}(2022)}]{gelderblom2022evaporation}%
  \BibitemOpen
  \bibfield  {author} {\bibinfo {author} {\bibfnamefont {H.}~\bibnamefont {Gelderblom}}, \bibinfo {author} {\bibfnamefont {C.}~\bibnamefont {Diddens}}, \ and\ \bibinfo {author} {\bibfnamefont {A.}~\bibnamefont {Marin}},\ }\bibfield  {title} {\enquote {\bibinfo {title} {Evaporation-driven liquid flow in sessile droplets},}\ }\href@noop {} {\bibfield  {journal} {\bibinfo  {journal} {Soft matter}\ } (\bibinfo {year} {2022})}\BibitemShut {NoStop}%
\bibitem [{\citenamefont {Jang}\ \emph {et~al.}(2023)\citenamefont {Jang}, \citenamefont {Yeom}, \citenamefont {Lee}, \citenamefont {Choi},\ and\ \citenamefont {Lee}}]{jang2023volatility}%
  \BibitemOpen
  \bibfield  {author} {\bibinfo {author} {\bibfnamefont {K.~H.}\ \bibnamefont {Jang}}, \bibinfo {author} {\bibfnamefont {S.~H.}\ \bibnamefont {Yeom}}, \bibinfo {author} {\bibfnamefont {H.~J.}\ \bibnamefont {Lee}}, \bibinfo {author} {\bibfnamefont {C.~K.}\ \bibnamefont {Choi}}, \ and\ \bibinfo {author} {\bibfnamefont {S.~H.}\ \bibnamefont {Lee}},\ }\bibfield  {title} {\enquote {\bibinfo {title} {Volatility effect on internal flow and contact line behavior during evaporation of binary mixture droplets},}\ }\href@noop {} {\bibfield  {journal} {\bibinfo  {journal} {International Journal of Heat and Mass Transfer}\ }\textbf {\bibinfo {volume} {207}},\ \bibinfo {pages} {124009} (\bibinfo {year} {2023})}\BibitemShut {NoStop}%
\bibitem [{\citenamefont {Hu}\ and\ \citenamefont {Larson}(2005{\natexlab{a}})}]{hu2005analysis}%
  \BibitemOpen
  \bibfield  {author} {\bibinfo {author} {\bibfnamefont {H.}~\bibnamefont {Hu}}\ and\ \bibinfo {author} {\bibfnamefont {R.~G.}\ \bibnamefont {Larson}},\ }\bibfield  {title} {\enquote {\bibinfo {title} {Analysis of the microfluid flow in an evaporating sessile droplet},}\ }\href@noop {} {\bibfield  {journal} {\bibinfo  {journal} {Langmuir}\ }\textbf {\bibinfo {volume} {21}},\ \bibinfo {pages} {3963--3971} (\bibinfo {year} {2005}{\natexlab{a}})}\BibitemShut {NoStop}%
\bibitem [{\citenamefont {Pan}\ \emph {et~al.}(2013)\citenamefont {Pan}, \citenamefont {Dash}, \citenamefont {Weibel},\ and\ \citenamefont {Garimella}}]{pan2013assessment}%
  \BibitemOpen
  \bibfield  {author} {\bibinfo {author} {\bibfnamefont {Z.}~\bibnamefont {Pan}}, \bibinfo {author} {\bibfnamefont {S.}~\bibnamefont {Dash}}, \bibinfo {author} {\bibfnamefont {J.~A.}\ \bibnamefont {Weibel}}, \ and\ \bibinfo {author} {\bibfnamefont {S.~V.}\ \bibnamefont {Garimella}},\ }\bibfield  {title} {\enquote {\bibinfo {title} {Assessment of water droplet evaporation mechanisms on hydrophobic and superhydrophobic substrates},}\ }\href@noop {} {\bibfield  {journal} {\bibinfo  {journal} {Langmuir}\ }\textbf {\bibinfo {volume} {29}},\ \bibinfo {pages} {15831--15841} (\bibinfo {year} {2013})}\BibitemShut {NoStop}%
\bibitem [{\citenamefont {Girard}\ \emph {et~al.}(2006)\citenamefont {Girard}, \citenamefont {Antoni}, \citenamefont {Faure},\ and\ \citenamefont {Steinchen}}]{girard2006evaporation}%
  \BibitemOpen
  \bibfield  {author} {\bibinfo {author} {\bibfnamefont {F.}~\bibnamefont {Girard}}, \bibinfo {author} {\bibfnamefont {M.}~\bibnamefont {Antoni}}, \bibinfo {author} {\bibfnamefont {S.}~\bibnamefont {Faure}}, \ and\ \bibinfo {author} {\bibfnamefont {A.}~\bibnamefont {Steinchen}},\ }\bibfield  {title} {\enquote {\bibinfo {title} {Evaporation and marangoni driven convection in small heated water droplets},}\ }\href@noop {} {\bibfield  {journal} {\bibinfo  {journal} {Langmuir}\ }\textbf {\bibinfo {volume} {22}},\ \bibinfo {pages} {11085--11091} (\bibinfo {year} {2006})}\BibitemShut {NoStop}%
\bibitem [{\citenamefont {Hu}\ and\ \citenamefont {Larson}(2006)}]{hu2006marangoni}%
  \BibitemOpen
  \bibfield  {author} {\bibinfo {author} {\bibfnamefont {H.}~\bibnamefont {Hu}}\ and\ \bibinfo {author} {\bibfnamefont {R.~G.}\ \bibnamefont {Larson}},\ }\bibfield  {title} {\enquote {\bibinfo {title} {Marangoni effect reverses coffee-ring depositions},}\ }\href@noop {} {\bibfield  {journal} {\bibinfo  {journal} {The Journal of Physical Chemistry B}\ }\textbf {\bibinfo {volume} {110}},\ \bibinfo {pages} {7090--7094} (\bibinfo {year} {2006})}\BibitemShut {NoStop}%
\bibitem [{\citenamefont {Hu}\ and\ \citenamefont {Larson}(2005{\natexlab{b}})}]{hu2005analysismarangoni}%
  \BibitemOpen
  \bibfield  {author} {\bibinfo {author} {\bibfnamefont {H.}~\bibnamefont {Hu}}\ and\ \bibinfo {author} {\bibfnamefont {R.~G.}\ \bibnamefont {Larson}},\ }\bibfield  {title} {\enquote {\bibinfo {title} {Analysis of the effects of marangoni stresses on the microflow in an evaporating sessile droplet},}\ }\href@noop {} {\bibfield  {journal} {\bibinfo  {journal} {Langmuir}\ }\textbf {\bibinfo {volume} {21}},\ \bibinfo {pages} {3972--3980} (\bibinfo {year} {2005}{\natexlab{b}})}\BibitemShut {NoStop}%
\bibitem [{\citenamefont {Lebedev}, \citenamefont {Silverman},\ and\ \citenamefont {Livhtenberg}(1965)}]{lebedev1965special}%
  \BibitemOpen
  \bibfield  {author} {\bibinfo {author} {\bibfnamefont {N.~N.}\ \bibnamefont {Lebedev}}, \bibinfo {author} {\bibfnamefont {R.~A.}\ \bibnamefont {Silverman}}, \ and\ \bibinfo {author} {\bibfnamefont {D.}~\bibnamefont {Livhtenberg}},\ }\href@noop {} {\enquote {\bibinfo {title} {Special functions and their applications},}\ } (\bibinfo {year} {1965})\BibitemShut {NoStop}%
\bibitem [{\citenamefont {Hu}\ and\ \citenamefont {Larson}(2002)}]{hu2002evaporation}%
  \BibitemOpen
  \bibfield  {author} {\bibinfo {author} {\bibfnamefont {H.}~\bibnamefont {Hu}}\ and\ \bibinfo {author} {\bibfnamefont {R.~G.}\ \bibnamefont {Larson}},\ }\bibfield  {title} {\enquote {\bibinfo {title} {Evaporation of a sessile droplet on a substrate},}\ }\href@noop {} {\bibfield  {journal} {\bibinfo  {journal} {The Journal of Physical Chemistry B}\ }\textbf {\bibinfo {volume} {106}},\ \bibinfo {pages} {1334--1344} (\bibinfo {year} {2002})}\BibitemShut {NoStop}%
\bibitem [{\citenamefont {Picknett}\ and\ \citenamefont {Bexon}(1977)}]{picknett1977evaporation}%
  \BibitemOpen
  \bibfield  {author} {\bibinfo {author} {\bibfnamefont {R.}~\bibnamefont {Picknett}}\ and\ \bibinfo {author} {\bibfnamefont {R.}~\bibnamefont {Bexon}},\ }\bibfield  {title} {\enquote {\bibinfo {title} {The evaporation of sessile or pendant drops in still air},}\ }\href@noop {} {\bibfield  {journal} {\bibinfo  {journal} {Journal of colloid and Interface Science}\ }\textbf {\bibinfo {volume} {61}},\ \bibinfo {pages} {336--350} (\bibinfo {year} {1977})}\BibitemShut {NoStop}%
\bibitem [{\citenamefont {Dunn}\ \emph {et~al.}(2009)\citenamefont {Dunn}, \citenamefont {Wilson}, \citenamefont {Duffy}, \citenamefont {David},\ and\ \citenamefont {Sefiane}}]{dunn2009strong}%
  \BibitemOpen
  \bibfield  {author} {\bibinfo {author} {\bibfnamefont {G.}~\bibnamefont {Dunn}}, \bibinfo {author} {\bibfnamefont {S.}~\bibnamefont {Wilson}}, \bibinfo {author} {\bibfnamefont {B.}~\bibnamefont {Duffy}}, \bibinfo {author} {\bibfnamefont {S.}~\bibnamefont {David}}, \ and\ \bibinfo {author} {\bibfnamefont {K.}~\bibnamefont {Sefiane}},\ }\bibfield  {title} {\enquote {\bibinfo {title} {The strong influence of substrate conductivity on droplet evaporation},}\ }\href@noop {} {\bibfield  {journal} {\bibinfo  {journal} {Journal of Fluid Mechanics}\ }\textbf {\bibinfo {volume} {623}},\ \bibinfo {pages} {329--351} (\bibinfo {year} {2009})}\BibitemShut {NoStop}%
\bibitem [{\citenamefont {Stauber}\ \emph {et~al.}(2014)\citenamefont {Stauber}, \citenamefont {Wilson}, \citenamefont {Duffy},\ and\ \citenamefont {Sefiane}}]{stauber2014lifetimes}%
  \BibitemOpen
  \bibfield  {author} {\bibinfo {author} {\bibfnamefont {J.~M.}\ \bibnamefont {Stauber}}, \bibinfo {author} {\bibfnamefont {S.~K.}\ \bibnamefont {Wilson}}, \bibinfo {author} {\bibfnamefont {B.~R.}\ \bibnamefont {Duffy}}, \ and\ \bibinfo {author} {\bibfnamefont {K.}~\bibnamefont {Sefiane}},\ }\bibfield  {title} {\enquote {\bibinfo {title} {On the lifetimes of evaporating droplets},}\ }\href@noop {} {\bibfield  {journal} {\bibinfo  {journal} {Journal of Fluid Mechanics}\ }\textbf {\bibinfo {volume} {744}},\ \bibinfo {pages} {R2} (\bibinfo {year} {2014})}\BibitemShut {NoStop}%
\bibitem [{\citenamefont {Wang}(2019)}]{wang2019contact}%
  \BibitemOpen
  \bibfield  {author} {\bibinfo {author} {\bibfnamefont {H.}~\bibnamefont {Wang}},\ }\bibfield  {title} {\enquote {\bibinfo {title} {From contact line structures to wetting dynamics},}\ }\href@noop {} {\bibfield  {journal} {\bibinfo  {journal} {Langmuir}\ }\textbf {\bibinfo {volume} {35}},\ \bibinfo {pages} {10233--10245} (\bibinfo {year} {2019})}\BibitemShut {NoStop}%
\bibitem [{\citenamefont {Shiri}\ \emph {et~al.}(2021)\citenamefont {Shiri}, \citenamefont {Sinha}, \citenamefont {Baumgartner},\ and\ \citenamefont {Cira}}]{shiri2021thermal}%
  \BibitemOpen
  \bibfield  {author} {\bibinfo {author} {\bibfnamefont {S.}~\bibnamefont {Shiri}}, \bibinfo {author} {\bibfnamefont {S.}~\bibnamefont {Sinha}}, \bibinfo {author} {\bibfnamefont {D.~A.}\ \bibnamefont {Baumgartner}}, \ and\ \bibinfo {author} {\bibfnamefont {N.~J.}\ \bibnamefont {Cira}},\ }\bibfield  {title} {\enquote {\bibinfo {title} {Thermal marangoni flow impacts the shape of single component volatile droplets on thin, completely wetting substrates},}\ }\href@noop {} {\bibfield  {journal} {\bibinfo  {journal} {Physical Review Letters}\ }\textbf {\bibinfo {volume} {127}},\ \bibinfo {pages} {024502} (\bibinfo {year} {2021})}\BibitemShut {NoStop}%
\bibitem [{\citenamefont {Kavehpour}, \citenamefont {Ovryn},\ and\ \citenamefont {McKinley}(2002)}]{kavehpour2002evaporatively}%
  \BibitemOpen
  \bibfield  {author} {\bibinfo {author} {\bibfnamefont {P.}~\bibnamefont {Kavehpour}}, \bibinfo {author} {\bibfnamefont {B.}~\bibnamefont {Ovryn}}, \ and\ \bibinfo {author} {\bibfnamefont {G.~H.}\ \bibnamefont {McKinley}},\ }\bibfield  {title} {\enquote {\bibinfo {title} {Evaporatively-driven marangoni instabilities of volatile liquid films spreading on thermally conductive substrates},}\ }\href@noop {} {\bibfield  {journal} {\bibinfo  {journal} {Colloids and Surfaces A: Physicochemical and Engineering Aspects}\ }\textbf {\bibinfo {volume} {206}},\ \bibinfo {pages} {409--423} (\bibinfo {year} {2002})}\BibitemShut {NoStop}%
\bibitem [{\citenamefont {Shen}, \citenamefont {Ren},\ and\ \citenamefont {Duan}(2020)}]{shen2020surface}%
  \BibitemOpen
  \bibfield  {author} {\bibinfo {author} {\bibfnamefont {L.}~\bibnamefont {Shen}}, \bibinfo {author} {\bibfnamefont {J.}~\bibnamefont {Ren}}, \ and\ \bibinfo {author} {\bibfnamefont {F.}~\bibnamefont {Duan}},\ }\bibfield  {title} {\enquote {\bibinfo {title} {Surface temperature transition of a controllable evaporating droplet},}\ }\href@noop {} {\bibfield  {journal} {\bibinfo  {journal} {Soft Matter}\ }\textbf {\bibinfo {volume} {16}},\ \bibinfo {pages} {9568--9577} (\bibinfo {year} {2020})}\BibitemShut {NoStop}%
\bibitem [{\citenamefont {Bird}, \citenamefont {Mandre},\ and\ \citenamefont {Stone}(2008)}]{bird2008short}%
  \BibitemOpen
  \bibfield  {author} {\bibinfo {author} {\bibfnamefont {J.~C.}\ \bibnamefont {Bird}}, \bibinfo {author} {\bibfnamefont {S.}~\bibnamefont {Mandre}}, \ and\ \bibinfo {author} {\bibfnamefont {H.~A.}\ \bibnamefont {Stone}},\ }\bibfield  {title} {\enquote {\bibinfo {title} {Short-time dynamics of partial wetting},}\ }\href@noop {} {\bibfield  {journal} {\bibinfo  {journal} {Physical review letters}\ }\textbf {\bibinfo {volume} {100}},\ \bibinfo {pages} {234501} (\bibinfo {year} {2008})}\BibitemShut {NoStop}%
\bibitem [{\citenamefont {Ding}\ and\ \citenamefont {Spelt}(2007)}]{ding2007inertial}%
  \BibitemOpen
  \bibfield  {author} {\bibinfo {author} {\bibfnamefont {H.}~\bibnamefont {Ding}}\ and\ \bibinfo {author} {\bibfnamefont {P.~D.}\ \bibnamefont {Spelt}},\ }\bibfield  {title} {\enquote {\bibinfo {title} {Inertial effects in droplet spreading: a comparison between diffuse-interface and level-set simulations},}\ }\href@noop {} {\bibfield  {journal} {\bibinfo  {journal} {Journal of fluid mechanics}\ }\textbf {\bibinfo {volume} {576}},\ \bibinfo {pages} {287--296} (\bibinfo {year} {2007})}\BibitemShut {NoStop}%
\bibitem [{\citenamefont {Chen}, \citenamefont {Bonaccurso},\ and\ \citenamefont {Shanahan}(2013)}]{chen2013inertial}%
  \BibitemOpen
  \bibfield  {author} {\bibinfo {author} {\bibfnamefont {L.}~\bibnamefont {Chen}}, \bibinfo {author} {\bibfnamefont {E.}~\bibnamefont {Bonaccurso}}, \ and\ \bibinfo {author} {\bibfnamefont {M.~E.}\ \bibnamefont {Shanahan}},\ }\bibfield  {title} {\enquote {\bibinfo {title} {Inertial to viscoelastic transition in early drop spreading on soft surfaces},}\ }\href@noop {} {\bibfield  {journal} {\bibinfo  {journal} {Langmuir}\ }\textbf {\bibinfo {volume} {29}},\ \bibinfo {pages} {1893--1898} (\bibinfo {year} {2013})}\BibitemShut {NoStop}%
\bibitem [{\citenamefont {Marin}\ \emph {et~al.}(2011)\citenamefont {Marin}, \citenamefont {Gelderblom}, \citenamefont {Lohse},\ and\ \citenamefont {Snoeijer}}]{marin2011order}%
  \BibitemOpen
  \bibfield  {author} {\bibinfo {author} {\bibfnamefont {A.~G.}\ \bibnamefont {Marin}}, \bibinfo {author} {\bibfnamefont {H.}~\bibnamefont {Gelderblom}}, \bibinfo {author} {\bibfnamefont {D.}~\bibnamefont {Lohse}}, \ and\ \bibinfo {author} {\bibfnamefont {J.~H.}\ \bibnamefont {Snoeijer}},\ }\bibfield  {title} {\enquote {\bibinfo {title} {Order-to-disorder transition in ring-shaped colloidal stains},}\ }\href@noop {} {\bibfield  {journal} {\bibinfo  {journal} {Physical review letters}\ }\textbf {\bibinfo {volume} {107}},\ \bibinfo {pages} {085502} (\bibinfo {year} {2011})}\BibitemShut {NoStop}%
\bibitem [{\citenamefont {Cazabat}\ and\ \citenamefont {Guena}(2010)}]{cazabat2010evaporation}%
  \BibitemOpen
  \bibfield  {author} {\bibinfo {author} {\bibfnamefont {A.-M.}\ \bibnamefont {Cazabat}}\ and\ \bibinfo {author} {\bibfnamefont {G.}~\bibnamefont {Guena}},\ }\bibfield  {title} {\enquote {\bibinfo {title} {Evaporation of macroscopic sessile droplets},}\ }\href@noop {} {\bibfield  {journal} {\bibinfo  {journal} {Soft Matter}\ }\textbf {\bibinfo {volume} {6}},\ \bibinfo {pages} {2591--2612} (\bibinfo {year} {2010})}\BibitemShut {NoStop}%
\bibitem [{\citenamefont {Wang}\ \emph {et~al.}(2023{\natexlab{a}})\citenamefont {Wang}, \citenamefont {Karapetsas}, \citenamefont {Valluri},\ and\ \citenamefont {Inoue}}]{wang2023spreading}%
  \BibitemOpen
  \bibfield  {author} {\bibinfo {author} {\bibfnamefont {Z.}~\bibnamefont {Wang}}, \bibinfo {author} {\bibfnamefont {G.}~\bibnamefont {Karapetsas}}, \bibinfo {author} {\bibfnamefont {P.}~\bibnamefont {Valluri}}, \ and\ \bibinfo {author} {\bibfnamefont {C.}~\bibnamefont {Inoue}},\ }\bibfield  {title} {\enquote {\bibinfo {title} {Spreading law of evaporative droplets},}\ }\href@noop {} {\bibfield  {journal} {\bibinfo  {journal} {arXiv preprint arXiv:2304.04030}\ } (\bibinfo {year} {2023}{\natexlab{a}})}\BibitemShut {NoStop}%
\bibitem [{\citenamefont {Wang}\ \emph {et~al.}(2023{\natexlab{b}})\citenamefont {Wang}, \citenamefont {Karapetsas}, \citenamefont {Valluri},\ and\ \citenamefont {Chihiro}}]{wang2023intricate}%
  \BibitemOpen
  \bibfield  {author} {\bibinfo {author} {\bibfnamefont {Z.}~\bibnamefont {Wang}}, \bibinfo {author} {\bibfnamefont {G.}~\bibnamefont {Karapetsas}}, \bibinfo {author} {\bibfnamefont {P.}~\bibnamefont {Valluri}}, \ and\ \bibinfo {author} {\bibfnamefont {I.}~\bibnamefont {Chihiro}},\ }\bibfield  {title} {\enquote {\bibinfo {title} {Intricate role of thermal properties and volatility in droplet spreading: A generalization to tanner’s law},}\ }\href@noop {} {\bibfield  {journal} {\bibinfo  {journal} {Bulletin of the American Physical Society}\ } (\bibinfo {year} {2023}{\natexlab{b}})}\BibitemShut {NoStop}%
\bibitem [{\citenamefont {Plesset}\ and\ \citenamefont {Prosperetti}(1976)}]{plesset1976flow}%
  \BibitemOpen
  \bibfield  {author} {\bibinfo {author} {\bibfnamefont {M.~S.}\ \bibnamefont {Plesset}}\ and\ \bibinfo {author} {\bibfnamefont {A.}~\bibnamefont {Prosperetti}},\ }\bibfield  {title} {\enquote {\bibinfo {title} {Flow of vapour in a liquid enclosure},}\ }\href@noop {} {\bibfield  {journal} {\bibinfo  {journal} {Journal of Fluid Mechanics}\ }\textbf {\bibinfo {volume} {78}},\ \bibinfo {pages} {433--444} (\bibinfo {year} {1976})}\BibitemShut {NoStop}%
\bibitem [{\citenamefont {Moosman}\ and\ \citenamefont {Homsy}(1980)}]{moosman1980evaporating}%
  \BibitemOpen
  \bibfield  {author} {\bibinfo {author} {\bibfnamefont {S.}~\bibnamefont {Moosman}}\ and\ \bibinfo {author} {\bibfnamefont {G.}~\bibnamefont {Homsy}},\ }\bibfield  {title} {\enquote {\bibinfo {title} {Evaporating menisci of wetting fluids},}\ }\href@noop {} {\bibfield  {journal} {\bibinfo  {journal} {Journal of Colloid and Interface Science}\ }\textbf {\bibinfo {volume} {73}},\ \bibinfo {pages} {212--223} (\bibinfo {year} {1980})}\BibitemShut {NoStop}%
\bibitem [{\citenamefont {Atkins}, \citenamefont {Atkins},\ and\ \citenamefont {de~Paula}(2014)}]{atkins2014atkins}%
  \BibitemOpen
  \bibfield  {author} {\bibinfo {author} {\bibfnamefont {P.}~\bibnamefont {Atkins}}, \bibinfo {author} {\bibfnamefont {P.~W.}\ \bibnamefont {Atkins}}, \ and\ \bibinfo {author} {\bibfnamefont {J.}~\bibnamefont {de~Paula}},\ }\href@noop {} {\emph {\bibinfo {title} {Atkins' physical chemistry}}}\ (\bibinfo  {publisher} {Oxford university press},\ \bibinfo {year} {2014})\BibitemShut {NoStop}%
\bibitem [{\citenamefont {Ajaev}(2005)}]{ajaev2005spreading}%
  \BibitemOpen
  \bibfield  {author} {\bibinfo {author} {\bibfnamefont {V.~S.}\ \bibnamefont {Ajaev}},\ }\bibfield  {title} {\enquote {\bibinfo {title} {Spreading of thin volatile liquid droplets on uniformly heated surfaces},}\ }\href@noop {} {\bibfield  {journal} {\bibinfo  {journal} {Journal of Fluid Mechanics}\ }\textbf {\bibinfo {volume} {528}},\ \bibinfo {pages} {279--296} (\bibinfo {year} {2005})}\BibitemShut {NoStop}%
\bibitem [{\citenamefont {Wang}\ \emph {et~al.}(2021)\citenamefont {Wang}, \citenamefont {Karapetsas}, \citenamefont {Valluri}, \citenamefont {Sefiane}, \citenamefont {Williams},\ and\ \citenamefont {Takata}}]{wang2021dynamics}%
  \BibitemOpen
  \bibfield  {author} {\bibinfo {author} {\bibfnamefont {Z.}~\bibnamefont {Wang}}, \bibinfo {author} {\bibfnamefont {G.}~\bibnamefont {Karapetsas}}, \bibinfo {author} {\bibfnamefont {P.}~\bibnamefont {Valluri}}, \bibinfo {author} {\bibfnamefont {K.}~\bibnamefont {Sefiane}}, \bibinfo {author} {\bibfnamefont {A.}~\bibnamefont {Williams}}, \ and\ \bibinfo {author} {\bibfnamefont {Y.}~\bibnamefont {Takata}},\ }\bibfield  {title} {\enquote {\bibinfo {title} {Dynamics of hygroscopic aqueous solution droplets undergoing evaporation or vapour absorption},}\ }\href@noop {} {\bibfield  {journal} {\bibinfo  {journal} {Journal of Fluid Mechanics}\ }\textbf {\bibinfo {volume} {912}},\ \bibinfo {pages} {A2} (\bibinfo {year} {2021})}\BibitemShut {NoStop}%
\bibitem [{\citenamefont {Xu}\ and\ \citenamefont {Luo}(2007)}]{xu2007marangoni}%
  \BibitemOpen
  \bibfield  {author} {\bibinfo {author} {\bibfnamefont {X.}~\bibnamefont {Xu}}\ and\ \bibinfo {author} {\bibfnamefont {J.}~\bibnamefont {Luo}},\ }\bibfield  {title} {\enquote {\bibinfo {title} {Marangoni flow in an evaporating water droplet},}\ }\href@noop {} {\bibfield  {journal} {\bibinfo  {journal} {Applied Physics Letters}\ }\textbf {\bibinfo {volume} {91}} (\bibinfo {year} {2007})}\BibitemShut {NoStop}%
\bibitem [{\citenamefont {Ristenpart}\ \emph {et~al.}(2007)\citenamefont {Ristenpart}, \citenamefont {Kim}, \citenamefont {Domingues}, \citenamefont {Wan},\ and\ \citenamefont {Stone}}]{ristenpart2007influence}%
  \BibitemOpen
  \bibfield  {author} {\bibinfo {author} {\bibfnamefont {W.}~\bibnamefont {Ristenpart}}, \bibinfo {author} {\bibfnamefont {P.}~\bibnamefont {Kim}}, \bibinfo {author} {\bibfnamefont {C.}~\bibnamefont {Domingues}}, \bibinfo {author} {\bibfnamefont {J.}~\bibnamefont {Wan}}, \ and\ \bibinfo {author} {\bibfnamefont {H.~A.}\ \bibnamefont {Stone}},\ }\bibfield  {title} {\enquote {\bibinfo {title} {Influence of substrate conductivity on circulation reversal in evaporating drops},}\ }\href@noop {} {\bibfield  {journal} {\bibinfo  {journal} {Physical review letters}\ }\textbf {\bibinfo {volume} {99}},\ \bibinfo {pages} {234502} (\bibinfo {year} {2007})}\BibitemShut {NoStop}%
\bibitem [{\citenamefont {Xu}, \citenamefont {Luo},\ and\ \citenamefont {Guo}(2010)}]{xu2010criterion}%
  \BibitemOpen
  \bibfield  {author} {\bibinfo {author} {\bibfnamefont {X.}~\bibnamefont {Xu}}, \bibinfo {author} {\bibfnamefont {J.}~\bibnamefont {Luo}}, \ and\ \bibinfo {author} {\bibfnamefont {D.}~\bibnamefont {Guo}},\ }\bibfield  {title} {\enquote {\bibinfo {title} {Criterion for reversal of thermal marangoni flow in drying drops},}\ }\href@noop {} {\bibfield  {journal} {\bibinfo  {journal} {Langmuir}\ }\textbf {\bibinfo {volume} {26}},\ \bibinfo {pages} {1918--1922} (\bibinfo {year} {2010})}\BibitemShut {NoStop}%
\bibitem [{\citenamefont {Parsa}\ \emph {et~al.}(2015)\citenamefont {Parsa}, \citenamefont {Harmand}, \citenamefont {Sefiane}, \citenamefont {Bigerelle},\ and\ \citenamefont {Deltombe}}]{parsa2015effect}%
  \BibitemOpen
  \bibfield  {author} {\bibinfo {author} {\bibfnamefont {M.}~\bibnamefont {Parsa}}, \bibinfo {author} {\bibfnamefont {S.}~\bibnamefont {Harmand}}, \bibinfo {author} {\bibfnamefont {K.}~\bibnamefont {Sefiane}}, \bibinfo {author} {\bibfnamefont {M.}~\bibnamefont {Bigerelle}}, \ and\ \bibinfo {author} {\bibfnamefont {R.}~\bibnamefont {Deltombe}},\ }\bibfield  {title} {\enquote {\bibinfo {title} {Effect of substrate temperature on pattern formation of nanoparticles from volatile drops},}\ }\href@noop {} {\bibfield  {journal} {\bibinfo  {journal} {Langmuir}\ }\textbf {\bibinfo {volume} {31}},\ \bibinfo {pages} {3354--3367} (\bibinfo {year} {2015})}\BibitemShut {NoStop}%
\bibitem [{\citenamefont {Girard}, \citenamefont {Antoni},\ and\ \citenamefont {Sefiane}(2008)}]{girard2008effect}%
  \BibitemOpen
  \bibfield  {author} {\bibinfo {author} {\bibfnamefont {F.}~\bibnamefont {Girard}}, \bibinfo {author} {\bibfnamefont {M.}~\bibnamefont {Antoni}}, \ and\ \bibinfo {author} {\bibfnamefont {K.}~\bibnamefont {Sefiane}},\ }\bibfield  {title} {\enquote {\bibinfo {title} {On the effect of marangoni flow on evaporation rates of heated water drops},}\ }\href@noop {} {\bibfield  {journal} {\bibinfo  {journal} {Langmuir}\ }\textbf {\bibinfo {volume} {24}},\ \bibinfo {pages} {9207--9210} (\bibinfo {year} {2008})}\BibitemShut {NoStop}%
\bibitem [{\citenamefont {Josyula}\ \emph {et~al.}(2018)\citenamefont {Josyula}, \citenamefont {Wang}, \citenamefont {Askounis}, \citenamefont {Orejon}, \citenamefont {Harish}, \citenamefont {Takata}, \citenamefont {Mahapatra},\ and\ \citenamefont {Pattamatta}}]{josyula2018evaporation}%
  \BibitemOpen
  \bibfield  {author} {\bibinfo {author} {\bibfnamefont {T.}~\bibnamefont {Josyula}}, \bibinfo {author} {\bibfnamefont {Z.}~\bibnamefont {Wang}}, \bibinfo {author} {\bibfnamefont {A.}~\bibnamefont {Askounis}}, \bibinfo {author} {\bibfnamefont {D.}~\bibnamefont {Orejon}}, \bibinfo {author} {\bibfnamefont {S.}~\bibnamefont {Harish}}, \bibinfo {author} {\bibfnamefont {Y.}~\bibnamefont {Takata}}, \bibinfo {author} {\bibfnamefont {P.~S.}\ \bibnamefont {Mahapatra}}, \ and\ \bibinfo {author} {\bibfnamefont {A.}~\bibnamefont {Pattamatta}},\ }\bibfield  {title} {\enquote {\bibinfo {title} {Evaporation kinetics of pure water drops: Thermal patterns, marangoni flow, and interfacial temperature difference},}\ }\href@noop {} {\bibfield  {journal} {\bibinfo  {journal} {Physical Review E}\ }\textbf {\bibinfo {volume} {98}},\ \bibinfo {pages} {052804} (\bibinfo {year} {2018})}\BibitemShut {NoStop}%
\bibitem [{\citenamefont {Li}\ \emph {et~al.}(2020)\citenamefont {Li}, \citenamefont {Diddens}, \citenamefont {Segers}, \citenamefont {Wijshoff}, \citenamefont {Versluis},\ and\ \citenamefont {Lohse}}]{li2020evaporating}%
  \BibitemOpen
  \bibfield  {author} {\bibinfo {author} {\bibfnamefont {Y.}~\bibnamefont {Li}}, \bibinfo {author} {\bibfnamefont {C.}~\bibnamefont {Diddens}}, \bibinfo {author} {\bibfnamefont {T.}~\bibnamefont {Segers}}, \bibinfo {author} {\bibfnamefont {H.}~\bibnamefont {Wijshoff}}, \bibinfo {author} {\bibfnamefont {M.}~\bibnamefont {Versluis}}, \ and\ \bibinfo {author} {\bibfnamefont {D.}~\bibnamefont {Lohse}},\ }\bibfield  {title} {\enquote {\bibinfo {title} {Evaporating droplets on oil-wetted surfaces: Suppression of the coffee-stain effect},}\ }\href@noop {} {\bibfield  {journal} {\bibinfo  {journal} {Proceedings of the National Academy of Sciences}\ }\textbf {\bibinfo {volume} {117}},\ \bibinfo {pages} {16756--16763} (\bibinfo {year} {2020})}\BibitemShut {NoStop}%
\bibitem [{\citenamefont {Zhang}\ \emph {et~al.}(2010)\citenamefont {Zhang}, \citenamefont {Li}, \citenamefont {Zhang},\ and\ \citenamefont {Yang}}]{zhang2010colloidal}%
  \BibitemOpen
  \bibfield  {author} {\bibinfo {author} {\bibfnamefont {J.}~\bibnamefont {Zhang}}, \bibinfo {author} {\bibfnamefont {Y.}~\bibnamefont {Li}}, \bibinfo {author} {\bibfnamefont {X.}~\bibnamefont {Zhang}}, \ and\ \bibinfo {author} {\bibfnamefont {B.}~\bibnamefont {Yang}},\ }\bibfield  {title} {\enquote {\bibinfo {title} {Colloidal self-assembly meets nanofabrication: From two-dimensional colloidal crystals to nanostructure arrays},}\ }\href@noop {} {\bibfield  {journal} {\bibinfo  {journal} {Advanced materials}\ }\textbf {\bibinfo {volume} {22}},\ \bibinfo {pages} {4249--4269} (\bibinfo {year} {2010})}\BibitemShut {NoStop}%
\bibitem [{\citenamefont {Park}\ and\ \citenamefont {Zhu}(2020)}]{park2020scalable}%
  \BibitemOpen
  \bibfield  {author} {\bibinfo {author} {\bibfnamefont {N.-G.}\ \bibnamefont {Park}}\ and\ \bibinfo {author} {\bibfnamefont {K.}~\bibnamefont {Zhu}},\ }\bibfield  {title} {\enquote {\bibinfo {title} {Scalable fabrication and coating methods for perovskite solar cells and solar modules},}\ }\href@noop {} {\bibfield  {journal} {\bibinfo  {journal} {Nature Reviews Materials}\ }\textbf {\bibinfo {volume} {5}},\ \bibinfo {pages} {333--350} (\bibinfo {year} {2020})}\BibitemShut {NoStop}%
\bibitem [{\citenamefont {Kim}\ \emph {et~al.}(2022)\citenamefont {Kim}, \citenamefont {Jo}, \citenamefont {Yun}, \citenamefont {Lee}, \citenamefont {Kim}, \citenamefont {Choi}, \citenamefont {Kang},\ and\ \citenamefont {Cho}}]{kim20223d}%
  \BibitemOpen
  \bibfield  {author} {\bibinfo {author} {\bibfnamefont {S.}~\bibnamefont {Kim}}, \bibinfo {author} {\bibfnamefont {S.~B.}\ \bibnamefont {Jo}}, \bibinfo {author} {\bibfnamefont {T.~K.}\ \bibnamefont {Yun}}, \bibinfo {author} {\bibfnamefont {S.}~\bibnamefont {Lee}}, \bibinfo {author} {\bibfnamefont {K.}~\bibnamefont {Kim}}, \bibinfo {author} {\bibfnamefont {Y.~J.}\ \bibnamefont {Choi}}, \bibinfo {author} {\bibfnamefont {J.}~\bibnamefont {Kang}}, \ and\ \bibinfo {author} {\bibfnamefont {J.~H.}\ \bibnamefont {Cho}},\ }\bibfield  {title} {\enquote {\bibinfo {title} {3d meniscus-guided evaporative assembly for rapid template-free synthesis of highly crystalline perovskite nanowire arrays},}\ }\href@noop {} {\bibfield  {journal} {\bibinfo  {journal} {Advanced Functional Materials}\ }\textbf {\bibinfo {volume} {32}},\ \bibinfo {pages} {2206264} (\bibinfo {year} {2022})}\BibitemShut {NoStop}%
\bibitem [{\citenamefont {Gao}\ \emph {et~al.}(2020)\citenamefont {Gao}, \citenamefont {Yan}, \citenamefont {Wang}, \citenamefont {Liu}, \citenamefont {Wu}, \citenamefont {Tang}, \citenamefont {Fang}, \citenamefont {Ding}, \citenamefont {Li}, \citenamefont {Sun} \emph {et~al.}}]{gao2020printable}%
  \BibitemOpen
  \bibfield  {author} {\bibinfo {author} {\bibfnamefont {A.}~\bibnamefont {Gao}}, \bibinfo {author} {\bibfnamefont {J.}~\bibnamefont {Yan}}, \bibinfo {author} {\bibfnamefont {Z.}~\bibnamefont {Wang}}, \bibinfo {author} {\bibfnamefont {P.}~\bibnamefont {Liu}}, \bibinfo {author} {\bibfnamefont {D.}~\bibnamefont {Wu}}, \bibinfo {author} {\bibfnamefont {X.}~\bibnamefont {Tang}}, \bibinfo {author} {\bibfnamefont {F.}~\bibnamefont {Fang}}, \bibinfo {author} {\bibfnamefont {S.}~\bibnamefont {Ding}}, \bibinfo {author} {\bibfnamefont {X.}~\bibnamefont {Li}}, \bibinfo {author} {\bibfnamefont {J.}~\bibnamefont {Sun}},  \emph {et~al.},\ }\bibfield  {title} {\enquote {\bibinfo {title} {Printable cspbbr 3 perovskite quantum dot ink for coffee ring-free fluorescent microarrays using inkjet printing},}\ }\href@noop {} {\bibfield  {journal} {\bibinfo  {journal} {Nanoscale}\ }\textbf {\bibinfo {volume} {12}},\ \bibinfo {pages} {2569--2577} (\bibinfo {year} {2020})}\BibitemShut {NoStop}%
\bibitem [{\citenamefont {Yu}\ \emph {et~al.}(2020)\citenamefont {Yu}, \citenamefont {Zhang}, \citenamefont {Long}, \citenamefont {Xu}, \citenamefont {Feng}, \citenamefont {Zhang}, \citenamefont {Liu},\ and\ \citenamefont {Xie}}]{yu2020efficient}%
  \BibitemOpen
  \bibfield  {author} {\bibinfo {author} {\bibfnamefont {H.}~\bibnamefont {Yu}}, \bibinfo {author} {\bibfnamefont {J.}~\bibnamefont {Zhang}}, \bibinfo {author} {\bibfnamefont {T.}~\bibnamefont {Long}}, \bibinfo {author} {\bibfnamefont {M.}~\bibnamefont {Xu}}, \bibinfo {author} {\bibfnamefont {H.}~\bibnamefont {Feng}}, \bibinfo {author} {\bibfnamefont {L.}~\bibnamefont {Zhang}}, \bibinfo {author} {\bibfnamefont {S.}~\bibnamefont {Liu}}, \ and\ \bibinfo {author} {\bibfnamefont {W.}~\bibnamefont {Xie}},\ }\bibfield  {title} {\enquote {\bibinfo {title} {Efficient all-blade-coated quantum dot light-emitting diodes through solvent engineering},}\ }\href@noop {} {\bibfield  {journal} {\bibinfo  {journal} {The Journal of Physical Chemistry Letters}\ }\textbf {\bibinfo {volume} {11}},\ \bibinfo {pages} {9019--9025} (\bibinfo {year} {2020})}\BibitemShut {NoStop}%
\end{thebibliography}%
\end{document}